\documentclass[11pt,letterpaper]{article} 

\usepackage[
  letterpaper,
  left=0.6in,right=0.6in,top=0.6in,bottom=0.6in,
  headheight=85pt,headsep=-45pt
]{geometry}
\usepackage{graphicx}
\graphicspath{{./}{figures/}}
\usepackage{float}
\usepackage{amsmath}
\usepackage[table]{xcolor}
\usepackage{tikz}

\usepackage{caption}
\usepackage{booktabs}
\usepackage{array}
\usepackage{tabularx}
\usepackage{enumitem}
\setlist{nosep}

\definecolor{tableheader}{HTML}{D9D9D9}
\newcolumntype{P}[1]{>{\sffamily\centering\arraybackslash}p{#1}}
\newcolumntype{Y}{>{\sffamily\centering\arraybackslash}X}
\newcolumntype{A}[1]{>{\raggedright\arraybackslash}p{#1}}
\newcolumntype{M}[1]{>{\raggedright\arraybackslash}m{#1}}

\usepackage[T1]{fontenc}
\usepackage[utf8]{inputenc}
\usepackage{PTSerif}
\usepackage[scaled]{helvet}

\newcommand{\headingfont}{\sffamily}

\usepackage{titlesec}
\titleformat{\section}
  {\headingfont\bfseries\raggedright\fontsize{18pt}{18pt}\selectfont}{}{0.75em}{}
\titleformat{\subsection}
  {\headingfont\bfseries\raggedright\fontsize{15pt}{16pt}\selectfont}{}{0.75em}{}
\titleformat{\subsubsection}
  {\headingfont\bfseries\raggedright\fontsize{12pt}{14pt}\selectfont}{}{0.75em}{}

\titlespacing*{\section}{0pt}{7pt}{6pt}
\titlespacing*{\subsection}{0pt}{7pt}{6pt}
\titlespacing*{\subsubsection}{0pt}{7pt}{6pt}

\newcommand{\miniheading}[1]{%
  \par\noindent{\headingfont\bfseries\fontsize{12pt}{14pt}\selectfont #1}\par\vspace{4pt}%
}

\usepackage{fancyhdr}
\fancyhfoffset[R]{18pt}
\makeatletter
\@ifundefined{theauthor}{}{}
\makeatother
\usepackage{titling}
\pretitle{\headingfont\bfseries\fontsize{24pt}{26pt}\selectfont\raggedright}
\posttitle{\par\vspace{-.3in}}
\preauthor{}\postauthor{}
\author{\mbox{}}
\date{}
\newcommand{\authorcard}[5]{%
  {\headingfont\bfseries\fontsize{12pt}{14pt}\selectfont #1}\par
  {\headingfont\bfseries\fontsize{12pt}{14pt}\selectfont #2}\par
  {\headingfont\bfseries\fontsize{12pt}{14pt}\selectfont #3}\par
  {\headingfont\bfseries\fontsize{12pt}{14pt}\selectfont #4}\par
  {\headingfont\bfseries\fontsize{12pt}{14pt}\selectfont #5}\par
}

\makeatletter
\newcommand{\authorpic}[1]{%
    \includegraphics[width=0.6in,height=0.6in,keepaspectratio,clip]{#1}%
}
\makeatother

\newcommand{\authorbioentry}[3]{%
  \noindent\begin{tabular}{@{}m{0.5in} M{\dimexpr\columnwidth-0.5in\relax}@{}}
    \authorpic{#1} & \textbf{#2}\par #3
  \end{tabular}\par\medskip
}

\makeatletter
\newcommand{\colfig}[2][]{%
  \IfFileExists{#2}{\includegraphics[width=\linewidth,#1]{#2}}{%
    \fbox{\parbox[b][1.5in][c]{\linewidth}{\centering \textit{Missing figure: }#2}}}%
}
\makeatother

\usepackage{csquotes}
\usepackage[style=apa,backend=biber]{biblatex}
\usepackage{changepage}

\usepackage{multicol}
\usepackage[hidelinks]{hyperref}
\newcommand{\mysubsection}[1]{\medskip\noindent\textbf{#1}}

\title{AI-Assisted Requirements Engineering:  An Empirical Evaluation Relative to Expert Judgment}

\begin{document}
\maketitle
\thispagestyle{empty}


\noindent
\begin{tabular*}{\textwidth}{@{\extracolsep{\fill}} A{0.32\textwidth} A{0.32\textwidth} A{0.32\textwidth}}
    \authorcard{Oz Levy}{Faculty of Industrial Engineering and Technology Management, Holon Institute of Technology (HIT)}{}{Holon, Israel}{ozl@my.hit.ac.il} &
  \authorcard{Ilya Dikman}{Faculty of Industrial Engineering and Technology Management, Holon Institute of Technology (HIT)}{}{Holon, Israel}{ilyad@my.hit.ac.il} &
  \authorcard{Natan Levy}{School of Computer Science and Engineering}{Hebrew University Jerusalem (HUJI)}{Jerusalem, Israel}{Natan.Levy1@mail.huji.ac.il} \\
  \multicolumn{3}{@{}c@{}}{\rule{0pt}{0.9\baselineskip}} \\[-0.2\baselineskip]
  \authorcard{Michael Winokur}{Faculty of Industrial Engineering and Technology Management, Holon Institute of Technology (HIT)}{}{Holon, Israel}{michaelw@hit.ac.il} &
\end{tabular*}
\addvspace{.75in}

\begin{multicols*}{2}
\raggedcolumns

\phantomsection
\miniheading{Abstract}
Artificial Intelligence (AI) is increasingly introduced into systems engineering activities, particularly within requirements engineering (RE), where quality assessment and validation remain heavily dependent on expert judgment. While recent AI tools demonstrate promising capabilities in analyzing and generating requirements, their role within formal systems engineering processes—and their alignment with established INCOSE criteria—remains insufficiently understood. This paper investigates the extent to which AI-based tools can support systems engineers in evaluating requirement quality, without replacing professional expertise.
The research adopts a structured systems engineering methodology to compare AI-assisted requirement evaluation with human expert assessment. A controlled study was conducted in which system requirements were evaluated against established INCOSE “good requirement” criteria by both experienced systems engineers and an AI-based assessment tool. The evaluation focused on consistency, completeness, clarity, and testability, examining not only accuracy but also the decision logic underlying each assessment.
Results indicate that AI tools can provide consistent and rapid preliminary assessments, particularly for syntactic and structural quality attributes. However, expert judgment remains essential for contextual interpretation, ambiguity resolution, and trade-off reasoning. Rather than positioning AI as a replacement for systems engineers, the findings support its role as a decision-support mechanism within the RE lifecycle. From a systems engineering perspective, this study contributes empirical evidence on how AI can be integrated into RE workflows while preserving traceability, accountability, and engineering consistency. The paper further discusses implications for SE practice, including workload reduction, quality assurance processes, and the boundaries of automation. The results inform both practitioners and researchers seeking to responsibly integrate AI into systems engineering processes.

\phantomsection
\subsubsection{Keywords}
Requirements engineering, large language models, requirement quality, functional and non-functional requirements, human-in-the-loop, AI-assisted engineering

\section{Introduction}
In recent years, AI has revolutionized various aspects of life and industry, becoming a key tool in processes once considered inapplicable to automated mechanisms (\cite{Stanford2024}). Leading organizations like Tesla and NASA have integrated AI into large-scale projects (\cite{TeslaAIPage}; \cite{nasa2016_seh}). For instance, Baidya (\cite{Baidya2022DigitalTwin}) explored the opportunities and challenges of using AI in critical projects like robotics and aeronautics, where AI assists with data analysis and provides insights into complex engineering tasks. 

In systems engineering, AI is expected to enhance the ability to analyze and classify engineering requirements, which form the foundation for product design and development. Existing research has explored AI’s role in classifying requirements in software and quality management systems (\cite{Cheligeer2022} ,\cite{TamaiAnzai2018}). However, these studies primarily cover specific, categorized tasks in software or quality requirements rather than providing a tool for comprehensive, standardized classification as per the International Council on Systems Engineering (INCOSE) standards (\cite{INCOSE2023}), which classify requirements into categories such as functional and non-functional.
With increasing demands for product quality and precision, analyzing requirements based on criteria like necessity, clarity, and verifiability is crucial. AI may enhance this process, offering rapid and accurate analyses that increase systems engineers’ efficiency. However, significant challenges remain due to AI’s tendency to generate misinterpretations, as seen in various applications (\cite{Maleki2024},\cite{Martinez2023}). Such inaccuracies could have severe consequences in complex engineering projects (\cite{Hadar2022},\cite{PerezCerrolaza2024}).
This study, therefore, proposes to evaluate the accuracy and reliability of AI models in such contexts, asking whether AI could match a systems engineer’s judgment or remains merely an auxiliary tool, similar to calculators. This research also examines the feasibility of integrating AI into engineering education, emphasizing how AI-based tools can equip engineers with 21st-century skills while fostering responsible engineering practices to address ethical challenges in critical systems. By bridging the gap between research and practical application, it aims to equip engineers with essential tools for today’s technological landscape.

\section{Background and Related Work}

Requirements engineering is a cornerstone of successful system development, providing a clear definition of what a system must do and under what conditions (\cite{Siddique2022}). High-quality requirements serve as the foundation for design and verification, whereas poorly defined requirements can lead to costly rework and project failures . Traditionally, requirements are documented in natural language due to its accessibility and legal familiarity in contracts . However, alternate media (such as models, diagrams, or prototypes) have been explored as complementary ways to express requirements, aiming to improve understanding and reduce ambiguity (\cite{Kolligs2025}). In parallel, the use of AI in RE has been growing. AI techniques have a long history in RE, from early knowledge-based systems to today’s data-driven approaches (\cite{Dalpiaz2020}). With modern AI breakthroughs – including powerful language models like GPT-4, Anthropic’s Claude (Sonnet series), and Meta’s LLaMA – there is renewed potential to automate and enhance many RE activities. 

\subsection{Engineering Requirements}
\subsubsection{Definition and Role in Systems Engineering}
In systems engineering  a requirement is typically a statement of needed functionality, constraint, or property that a system must satisfy to meet stakeholder objectives (\cite{Siddique2022b}). Requirements engineering involves eliciting these needs and translating them into precise system specifications (\cite{Siddique2022b}). Well-written requirements ensure that development teams and stakeholders share a common understanding of the system’s goals.

\begin{figure}[H]
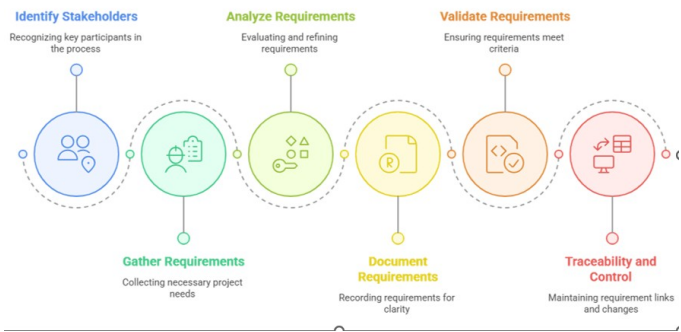

  \centering
  \colfig{figures/fig1.jpg}
  \caption{Requirements Engineering Process: A Flowchart of Core Stages.}
  \label{fig:fig1}
\end{figure}

 Conversely, inadequate requirements are often cited as a primary cause of project issues – “low quality requirements are among the first potential mistakes… felt downstream in the form of increased costs and schedule overruns” (\cite{Kolligs2025}). Clear and precise requirements make it easier to verify the system and avoid costly late-stage changes (\cite{INCOSE2023}).

\subsection{Classification of Requirements}
Requirements can be classified along several dimensions. A common classification is functional vs. non-functional requirements. Functional requirements specify the services, behaviors, or functions a system must execute – essentially, what the system should do. Non-functional requirements (NFRs), in contrast, define how the system should perform or the qualities it must have (\cite{Siddique2022b}). These include performance metrics, security levels, usability, reliability, and other quality attributes or constraints. For example, an online service’s functional requirement might be the ability to register a new user, while a non-functional requirement could stipulate that the registration response occurs within 2 seconds (a performance constraint). Both types are vital: functional requirements ensure the system’s features meet user needs, and NFRs ensure the system is usable and trustworthy under real-world conditions (\cite{Siddique2022b}). 
\textbf{\textit{}}

\subsection{Distinguishing Effective from Deficient Requirements}

The distinction between effective and deficient requirements in this study follows the quality characteristics defined in the INCOSE \emph{Guide to Writing Requirements} (GtWR) (\cite{INCOSE2023}). Rather than redefining these principles here, the paper adopts the operational definition and evaluation criteria described in the Methodology section below, where requirement quality is assessed using seven INCOSE-aligned attributes suitable for objective human and AI evaluation (\cite{INCOSE2023}).

\subsection{Language Models}
\mysubsection{Historical Evolution and Breakthroughs in AI}
Artificial intelligence has progressed through three main waves, each reshaping how RE activities are automated and supported.
\begin{figure}[H]
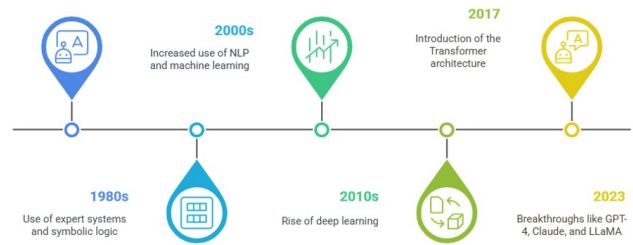

  \centering
  \colfig{figures/fig2.jpg}
  \caption{The Evolution of Artificial Intelligence in Requirements Engineering: A Chronological Perspective.}
  \label{fig:fig2}
\end{figure}

\begin{enumerate}
    \item{Symbolic AI (mid-1950s – early 1990s).}
 The first wave focused on symbolic reasoning, in which expert systems captured domain knowledge with IF–THEN rules and formal logic. During the 1980s, such rule-based engines were already inspecting requirements documents for internal consistency (\cite{DalpiazNiu2020}). Although effective for narrow domains, these systems struggled with ambiguous natural language and quickly became brittle as project scope grew.
 \item{Data-Driven Machine Learning (1990s – late 2000s).}
 Rising computational power enabled statistical natural-language processing and classical machine-learning (ML) algorithms for example, support-vector machines and decision trees. Researchers began to prioritize, classify and cluster large sets of requirements, reducing manual workload while highlighting potential defects (\cite{DalpiazNiu2020}). At the same time, automated detectors of vague or subjective terms emerged, flagging words such as fast or user-friendly that undermine precision in software specifications (\cite{Kolligs2025}).
\item{Deep Learning and Large Language Models (2010s – present).}
 The third wave was triggered by multilayer neural networks and, above all, the Transformer architecture (\cite{Vaswani2017}). Deep models trained on vast corpora achieved state-of-the-art results across computer vision, speech and NLP. Transfer learning allowed researchers to fine-tune pre-trained language models (e.g., BERT) on modest RE datasets, yielding substantial gains in tasks such as functional/non-functional classification (\cite{Kaur2024}). Martínez-Fernández(\cite{MartinezFernandez2022}) describes this period as the “rising wave of deep learning,” characterized by increasingly context-aware techniques that can parse entire specification documents, map traceability links and even generate draft requirements.

\end{enumerate}
Together, these three waves chart a shift from manually encoded rules to data-centric, context-sensitive analysis transforming RE from labour-intensive bookkeeping into a semi-automated, insight-driven discipline.
\subsection{Artificial Intelligence in Requirements Engineering}

AI has been increasingly applied to support RE, primarily through natural language processing (NLP) and machine learning techniques aimed at reducing manual effort and improving consistency. Early work focused on narrow, task-specific automation such as classifying requirements as functional or non-functional, detecting vague or subjective language, or clustering requirements by topic. Systematic reviews show that, given sufficient training data, machine-learning models can achieve high accuracy on well-defined classification tasks, particularly in software requirements datasets (e.g., \cite{Dalpiaz2020}, \cite{Kaur2024}).

Recent advances in large language models (LLMs) have expanded these capabilities. Transformer-based models can process longer contexts and capture semantic intent beyond surface-level keywords, enabling more nuanced classification and annotation of requirements. Systems such as NLP4ReF demonstrate that LLM-based approaches can outperform traditional NLP pipelines in requirement classification and even generate candidate requirements at low cost and high speed (\cite{Peer2024}). These results indicate that AI can meaningfully assist with routine RE activities, especially in large-scale or time-constrained projects.

However, existing research remains largely fragmented and domain-specific. Most studies focus on isolated tasks (e.g., FR/NFR classification or requirement generation), are grounded in software-centric datasets, or rely on accuracy metrics without comparing AI behavior to professional systems-engineering judgment. Moreover, while several approaches implicitly reference quality attributes, few explicitly align AI evaluations with the structured criteria defined in the INCOSE Guide to Writing Requirements. As a result, it remains unclear to what extent AI assessments correspond to how experienced systems engineers reason about requirement quality, ambiguity, and acceptability.

Another limitation concerns explainability and accountability. Although prompt engineering, expert personas, and structured outputs have been shown to improve AI performance and usability, AI systems still lack access to technical truth, system context, and cross-requirement reasoning. Consequently, the literature consistently emphasizes that AI should function as a decision-support tool rather than an autonomous evaluator, with human experts retaining responsibility for validation and trade-off decisions (\cite{Bender2021}, \cite{Shneiderman2020}).

Taken together, prior work suggests strong potential for AI to support RE, but also reveals a gap between task-level automation and engineering-grade evaluation aligned with INCOSE practice. Addressing this gap requires empirical comparison between AI-based assessments and human expert judgment, using clearly defined quality criteria and controlled experimental design. The following section presents the methodology adopted in this study to evaluate AI performance against experienced systems engineers across both requirement quality assessment and functional/non-functional classification.

\section{Methodology}
\begin{figure}[H]
  \centering
  \colfig{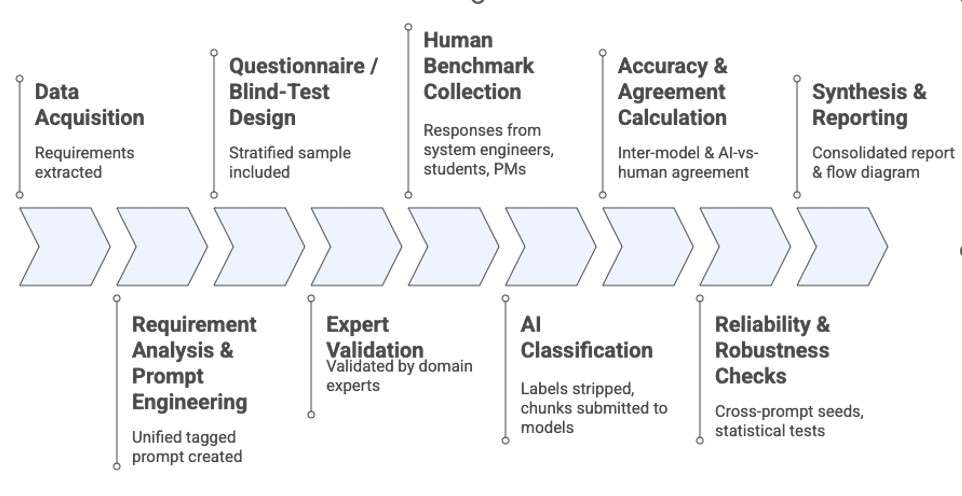}
  \caption{The Journey of AI Research and Development: From Requirements Acquisition to Findings Reporting}
  \label{Fig:Fig5}
\end{figure}

This section explains how the research was designed and carried out. The goal was to evaluate how well AI can understand, classify, and assess engineering requirements compared to human experts. The methodology combines traditional systems engineering practices with modern AI techniques, creating a structured process from data collection to result analysis.
Two main datasets were used, one from a real-world case study (Dr. Tools )(\cite{Hadar2022}) and another public dataset (PROMISE)(\cite{ClelandHuang2006}) to test AI performance in different contexts. Each step of the process, from requirement review and prompt design to model execution and human comparison, was carefully planned to ensure reliable and meaningful results. The following subsections describe each stage of the methodology in detail.
Figure \ref{Fig:Fig5} presents the complete process of the AI research and development journey, from collecting requirements to reporting results. The workflow includes several key stages: acquiring and analyzing data, designing prompts, validating inputs with experts, and performing AI-based classification. Human benchmarks are then collected to compare AI performance with expert evaluations, followed by accuracy, agreement, and robustness checks to ensure reliability. The process concludes with synthesis and reporting, where all findings are integrated into a comprehensive summary of results.

This adaptation preserves the conceptual integrity of INCOSE while enabling a standardized, AI-friendly evaluation framework.
The methodology addresses three research questions:

\begin{enumerate}
    \item \emph{RQ1:} To what extent can AI classify engineering requirements according to INCOSE “good requirement” criteria compared to an experienced systems engineer?
    \item \emph{RQ2:} How effectively can AI distinguish between functional and non-functional requirements compared to an experienced systems engineer?	
    \item \emph{RQ3:} What are the advantages and limitations of AI compared to human experts in understanding and classifying engineering requirements?
\end{enumerate}

\subsection{Data Collection}
\begin{figure}[H]
  \centering
  \colfig{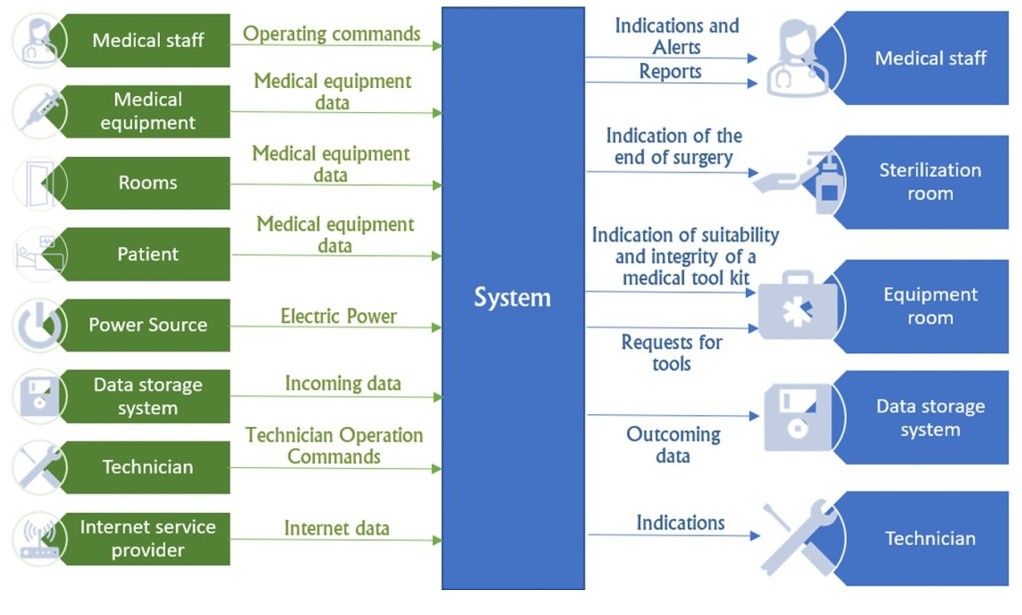}
  \caption{Context Diagram of Dr. Tools system.}
  \label{Fig:Fig6}
\end{figure}

Two independent datasets were used:
Project A – DR Tool Case Study:	
Originating from a graduate engineering project at the Holon Institute of Technology, this dataset involves an RFID-enabled inventory management system for operating rooms. It includes 31 stakeholder and 76 optimized system requirements, covering:

\begin{enumerate}
\item Functional goals: real-time instrument localization, set verification before incision, and automated alerts.	
\item Non-functional goals: reliability, security, and integration with hospital IT systems.
\end{enumerate}
The context diagram (Figure \ref{Fig:Fig6}) illustrates the system’s environment, highlighting interactions among surgical staff, sterile-processing departments, and hospital databases. This rich requirement set serves as a benchmark for evaluating AI classification against expert systems engineering practice.

Project B – PROMISE Dataset:
This publicly available corpus contains 969 software requirements (444 functional, 525 non-functional), distributed across twelve non-functional sub-classes (e.g., Security, Usability, Performance). To address class imbalance, SMOTE–Tomek resampling was applied within a stratified ten-fold cross-validation pipeline. Logistic regression achieved 76.16 \% ± 2.58 \% accuracy, a significant improvement over the 58.31 \% ± 2.05 \% baseline, underscoring the value of balanced training data.

\begin{figure}[H]
  \centering
  \colfig{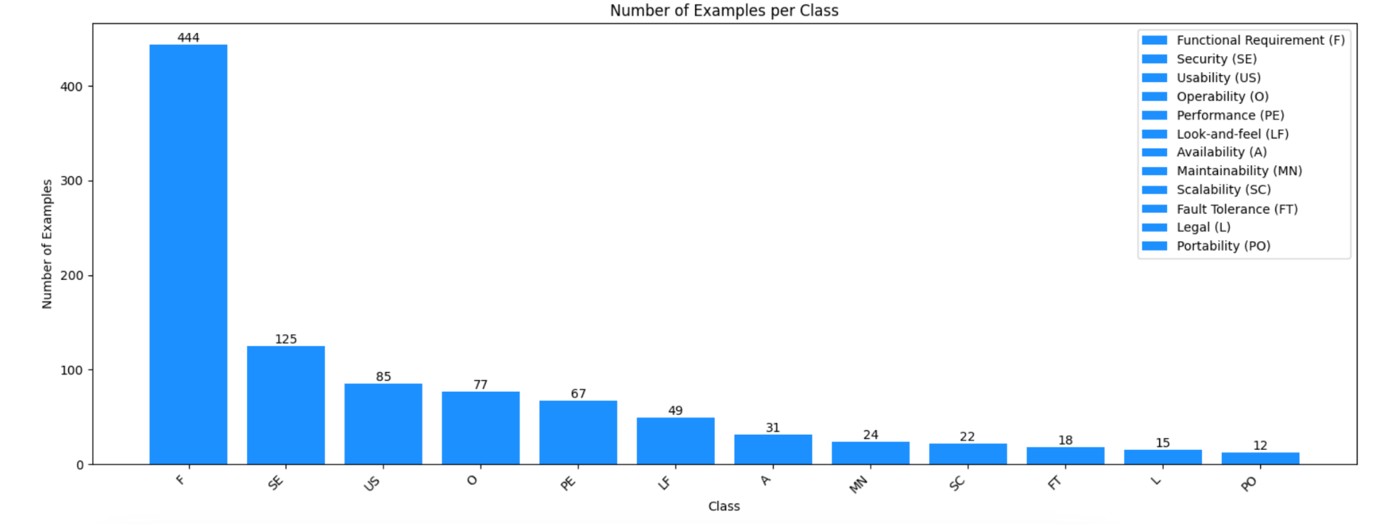}
  \caption{Requirement type distribution in the PROMISE dataset.}
  \label{Fig:Fig7}
\end{figure}

The bar chart in Figure \ref{Fig:Fig7}, titled “Number of Examples per Class,” presents the distribution of requirement examples across different categories. The data show that Functional Requirements (F) dominate the dataset with 444 examples, significantly more than any other class. Following this, Security (SE) has 125 examples, Usability (US) 85, Operability (O) 77, and Performance (P) 67. The remaining classes such as Look-and-Feel (LF), Availability (A), Maintainability (MN), Scalability (SC), Fault Tolerance (FT), Legal (L), and Portability (PO) contain relatively few examples, ranging from 49 down to 17. This indicates a strong imbalance in the dataset, with functional requirements being the most represented category by a large margin.

\subsection{How the Requirements Were Reviewed}
The process of reviewing the requirements aimed to ensure that each statement met recognized quality standards and could be effectively analyzed by the proposed AI framework. To achieve this, the study followed the principles outlined in the INCOSE Guide to Writing Requirements (GtWR), which defines the characteristics of a “good requirement.” These principles served as both the foundation for human review and the benchmark for automated evaluation.

According to INCOSE, a good requirement must be clear, precise, and feasible. Building on this foundation, the present research examined how Natural Language Processing (NLP) and Machine Learning (ML) techniques can assist in automatically analyzing and classifying requirements. The proposed AI pipeline integrates text classification, clustering, and topic modeling using models such as Support Vector Machines (SVM), Convolutional Neural Networks (CNNs), and Latent Dirichlet Allocation (LDA).

To differentiate between functional and non-functional requirements, the Quality Requirements Mining and Classification Process was applied, combining Word2Vec and Doc2Vec vectorization methods with CNN architectures. This enables large-scale analysis of Software Requirements Specifications (SRS) in accordance with ISO/IEC 25030 standards. Additionally, the methodology includes AI-assisted test case generation, where LSTM and Transformer-based models extract contextual information to automatically produce preliminary test cases
.
As part of the review framework, INCOSE’s nine quality traits were used as the reference point, with terminology adapted to align with ISO/IEC/IEEE 29148:2018, IEEE 830:1998, and the NASA Systems Engineering Handbook (2016). The traits Necessary and Appropriate were replaced with Essential and Independent, emphasizing the need for requirements to be indispensable and solution neutral. The terms Correct and Conforming were excluded from automated assessment, as they require domain-specific and organizational context not yet measurable by AI.
Thus, the review process focused on seven key attributes that could be objectively evaluated: Essential, Independent, Unambiguous, Complete, Singular, Feasible, and Verifiable ensuring that each requirement was both human-validated and machine-assessable according to international quality standards.
Two characteristics “Correct” and “Conforming” were intentionally omitted from the automated evaluation because:
\begin{enumerate}
    \item Correct requires domain-specific verification against underlying analysis.
    \item Conforming depends on organization-specific style and pattern guidelines.
\end{enumerate}

Since these attributes cannot be reliably evaluated through current AI/NLP methods, the research focuses on the seven characteristics that can be assessed objectively:

\begin{enumerate}
    \item \emph{Essential:} Defines a necessary capability or constraint.
    \item \emph{Independent:} Specifies what is needed, not how to implement it. 
    \item \emph{Unambiguous:} Can be interpreted in only one way.
    \item \emph{Singular:} Expresses a single idea.
    \item \emph{Feasible:} Achievable within project constraints.
    \item \emph{Verifiable:} Can be validated through inspection, analysis, or testing. 
\end{enumerate}

For the DR Tool, three certified systems engineers independently reviewed each requirement, classifying it as either “good” or “not good.” For requirements deemed not good, they also documented the specific reasons for the quality issues. The reviewers then held a consensus meeting to resolve discrepancies and agree on a final classification. This review and consensus process aligns with the evaluation approach described in Section 3.5.
For PROMISE, existing expert labels were validated by a single expert reviewer before use, in accordance with the validation process outlined in Section 3.6.

\subsection{Prompt Engineering}
Two separate prompt templates were developed:
\begin{itemize}
    \item \emph{Experiment 1 – DR Tool:} The prompt instructed the model to evaluate each requirement against the seven INCOSE criteria and explain which were met or violated. The goal was to assess requirement quality. [Appendix B].
    \item \emph{Experiment 2 – PROMISE:} The prompt instructed the model to classify each requirement as Functional or Non-Functional and, for the latter, specify the appropriate NF sub-category.  
\end{itemize}

Together, these experiments provide complementary insights: quality evaluation (DR Tool) vs. type classification (PROMISE).

\subsection{Model Execution}
Model Execution
All data were processed in batches of $\leq$ 3,000 tokens using PromptTag. Each batch was evaluated by ChatGPT-4, Claude Sonnet 3.5, and Meta Llama 3. Raw outputs were stored verbatim for subsequent analysis.

\subsection{Human Survey (Project A)}
To benchmark AI performance against human judgment, a Google Forms survey was distributed to over 300 systems engineers worldwide. A set of 20 requirements selected based on model disagreement was used. Participants (n = 21 complete, 2 partial) performed two tasks per requirement:

\begin{enumerate}
    \item Indicate whether the requirement is acceptable or problematic.
    \item Identify which of the seven INCOSE - like criteria are violated.
\end{enumerate}

The survey was pilot tested with a small group of participants to identify potential issues and refine question clarity before being distributed to the broader user base. Results from a two-layer human baseline: overall acceptability and specific quality violations [Appendix A].

\subsection{AI-Only Evaluation (Project B)}
PROMISE requirements were classified by the three models without human intervention. Each model assigned a label (F/NF) and, where relevant, an NF sub-class (e.g., Reliability, Security). Predictions were compared against reference labels. 

\subsection{Comparative Analysis}
Performance metrics were computed in Python 3.11 for each model and dataset:
\begin{itemize}
    \item Accuracy (quality classification for DR Tool; F/NF for PROMISE)
    \item Inter-model agreement 
    \item Precision, recall, and $F_1$ scores (against the human baseline for Project A)
\end{itemize}

Areas of disagreement between AI and humans were analyzed to reveal systematic ambiguity patterns.

\subsection{Synthesis and Reporting}
The results from AI evaluations, expert review, and human surveys were triangulated to ensure methodological Validity. A detailed flow diagram outlines inputs, processes, validation gates, and outputs, leading to actionable recommendations for operationalizing AI in requirements engineering.

\section{Results}
This chapter presents the outcomes of applying the methodology described in Chapter 3. The evaluation consisted of two complementary analyses. First, the ability of a LLM to identify poorly defined system engineering requirements according to the INCOSE Guide to Writing Requirements was compared against survey results provided by professional engineers. 

Second, the LLM’s capability to classify software requirements into functional 
and non-functional categories was assessed against the pre-labeled PROMISE\_exp dataset.

\subsection{AI Model Performance in Requirements Classification}
Section 4.1 presents an in-depth empirical analysis that goes beyond traditional consensus-based evaluation to examine individual engineer assessments across the complete dataset. Rather than simply comparing AI models against a single consensus outcome, we analyze 420 individual assessment points (21 engineers × 20 requirements) to provide a more nuanced understanding of AI model performance in requirements quality assessment.

\mysubsection{Introduction and Expanded Methodology.}
The expanded methodology examines how three state-of-the-art AI language models Claude Sonnet 3.5, GPT-4, and Llama 3 perform when compared against each individual engineer's assessment, revealing patterns of agreement and disagreement that are obscured by consensus-only analysis. This granular approach provides insights into the variability of human expert judgment and the consistency of AI assessments across different evaluation perspectives.

\mysubsection{Dataset Overview and Expert Panel Statistics.}
\begin{itemize}
    \item Total Engineers: 21 systems engineering professionals
    \item Requirements Evaluated: 20 medical equipment tracking system requirements
    \item Individual Assessment Points: 420 ($21 \times 20$)
    \item Valid Responses: 380 (19 engineers $\times$ 20 requirements due to some missing responses)
    \item Experience Distribution: 86\% with 8+ years, 14\% with less than 8 years
    \item AI Models Tested: Claude Sonnet 3.5, GPT-4, Llama 3
\end{itemize}

\begin{figure}[H]
  \centering
  \colfig{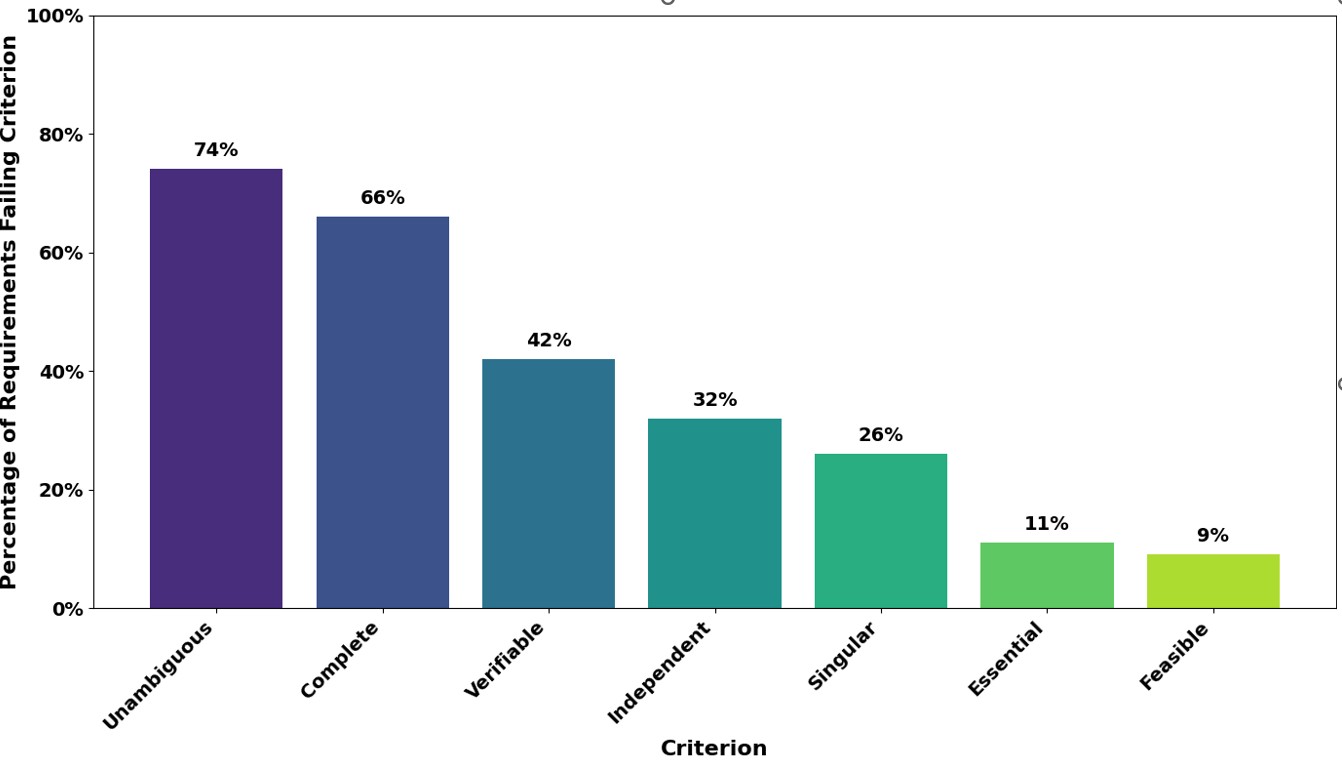}
  \caption{INCOSE Quality Criteria Violations.}
  \label{Fig:Fig8}
\end{figure}

Figure \ref{Fig:Fig8} presents the distribution of requirements as classified by the engineers who responded to the survey, based on the quality criteria that they fail to meet. Each requirement may be considered non-compliant with more than one criterion, and the percentages represent the proportion of requirements that fail to meet the standards for each criterion. 74\% of the requirements were found to be unambiguous , 66\% fail the Complete criterion, and 42\% Verifiable criterion. Furthermore, 32\% Independent criterion, 26\% Singular criterion (i.e., they express more than one idea), and when looking at the low percentages, 11\% Essential criterion and 9\% Feasible criterion (i.e., they are not achievable). This data indicates that clarity is strong, but practicality and testability are limited.

\mysubsection{Individual Engineer Assessment Patterns.}
Analysis of individual engineer assessments reveals significant variability in quality evaluation stringency. While some engineers identified issues in nearly all requirements, others were more lenient in their assessments. This variability provides a crucial context for understanding AI model performance.

\begin{table*}[ht]
\centering
\begin{tabular}{||l|c|c|c|c||}

\hline
Req. ID & Engineers Finding Issues & Total Responses & Percentage & Consensus \\
\hline
\hline
RS1.1.1  & 16 & 19 & 84\% & Not Good \\
RS2.0.1  & 14 & 19 & 74\% & Not Good \\
RS3.0.1  & 13 & 19 & 68\% & Not Good \\
RS5.1.2  & 10 & 19 & 53\% & Not Good \\
RS6.0.1  & 12 & 19 & 63\% & Not Good \\
RS9.1.1  & 15 & 20 & 75\% & Not Good \\
RS11.0.1 & 16 & 19 & 84\% & Not Good \\
RS14.0.1 & 16 & 20 & 80\% & Not Good \\
RS19.2.3 & 11 & 20 & 55\% & Not Good \\
RS19.2.4 & 12 & 20 & 60\% & Not Good \\
RS8.0.1  & 13 & 20 & 65\% & Not Good \\
RS11.0.2 & 14 & 20 & 70\% & Not Good \\
RS24.1.1 & 11 & 20 & 55\% & Not Good \\
RS18.2.2 & 15 & 20 & 75\% & Not Good \\
RS1.2.1  & 14 & 20 & 70\% & Not Good \\
RS9.1.2  & 16 & 20 & 80\% & Not Good \\
RS15.0.1 & 13 & 20 & 65\% & Not Good \\
RS17.0.1 & 8  & 20 & 40\% & Disputed \\
RS19.1.1 & 2  & 20 & 10\% & Good \\
RS19.2.1 & 12 & 20 & 60\% & Not Good \\
\hline
\hline
\end{tabular}
\caption{Engineers findings per requirement}
\label{tab:req_findings}
\end{table*}

Engineer agreement varies significantly across requirements, ranging from 10\% (RS19.1.1) to 84\% (RS1.1.1 and RS11.0.1) finding quality issues. This variation indicates that some requirements have obvious defects while others are subject to interpretation differences among experts.

\mysubsection{AI Model Performance Against Individual Engineers.}
When comparing AI models against individual engineer assessments rather than just consensus, a more complex performance picture emerges. We calculate accuracy for each AI model against all 380 individual assessment points, providing a statistically robust evaluation of model performance.

\mysubsection{Detailed Performance Metrics.}

\begin{table*}[ht]
\centering
\begin{tabular}{||l|c|c|c|c||}
\hline
\textbf{Model} & \textbf{Point Estimate} & \textbf{95\% CI Lower} & \textbf{95\% CI Upper} & \textbf{Margin of Error} \\
\hline
\hline
Claude Sonnet 3.5 & 85.0\% & 81.2\% & 88.3\% & 3.55\% \\
GPT-4o            & 45.0\% & 40.1\% & 49.8\% & 4.85\% \\
Llama 3           & 47.9\% & 42.8\% & 53.1\% & 5.15\% \\
\hline
\hline
\end{tabular}
\caption{Performance Confidence Intervals (95\% CI) via Bootstrap Resampling ($n=10,000$)}
\label{tab:performance_ci}
\end{table*}

Table~\ref{tab:model_agreement} summarizes how often each model produced the same answer as the engineers.

\begin{itemize}
    \item \emph{Claude Sonnet 3.5} achieved the strongest performance, matching human judgments in 85\% of the comparisons.
    \item \emph{GPT-4} and \emph{Llama 3} showed substantially lower agreement, at approximately 45-48\%.
\end{itemize}

The final column, \textit{Standard Deviation (Std. Deviation)}, indicates how consistent each model was across different engineers:

\begin{itemize}
    \item A smaller value (e.g., $\pm12.3\%$) indicates stable behavior, meaning the model performed similarly with all evaluators.
    \item A larger value (around $\pm18\%$) suggests the model's accuracy varied significantly, agreeing well with some engineers but poorly with others.
\end{itemize}

In summary, \emph{Claude Sonnet 3.5} was both the most accurate and the most consistent model.  
In contrast, \emph{GPT-4} and \emph{Llama 3} demonstrated lower accuracy and greater variability across individual evaluators.

\mysubsection{Key Findings and Implications.}
The key findings reveal strong differentiation in model performance and human consistency. When compared against 380 individual engineer evaluations, Claude Sonnet 3.5 achieved an impressive 85\% accuracy, significantly outperforming GPT-4 (45\%) and Llama 3 (47.9\%). Claude also showed the highest consistency across evaluators, with a performance range of only 25\%, while GPT-4 and Llama 3 fluctuated more widely (range: 35\%). These performance gaps are statistically significant, as demonstrated by the non-overlapping 95\% confidence intervals derived via bootstrap resampling with 10,000 iterations. 

\subsection{Classification of Functional and Non-Functional Software Requirements  (LLM vs. PROMISE\_exp Dataset)}
The PROMISE\_exp dataset comprises 969 software requirements, of which 444 are functional and 525 are non-functional. 
The non-functional requirements are further divided into eleven subcategories: Quality (SE, 125), Usability (US, 85), Interface (O, 77), Performance (PE, 67), Legal and Finance (LF, 49), Availability (A, 31), Maintainability (MN, 24), Security (SC, 22), Fault Tolerance (FT, 18), and Portability (L, 15 $\&$ PO, 12). Three large language models Claude 3.5 Sonnet, GPT-4o, and Llama 3.0 were tasked first with distinguishing functional vs. non-functional requirements and then assigning non-functional items to the correct subcategory. Model outputs were compared against the ground-truth labels to assess per-category accuracy, identify best performers, and quantify performance gaps. NASA and INCOSE do not explicitly prescribe a fixed taxonomy for non-functional requirements (NFRs). In practice, several frameworks are used, and in Israel, a common approach is to map PROMISE dataset codes to standard Systems Engineering (SE) NFR categories.

Table \ref{tab:promise_se_nfr_mapping} presents a mapping that enables a unified classification across projects,  applied in our working environment as stated above,  facilitating comparative analysis and traceability.
\begin{table*}[ht]
\centering
\begin{tabular}{||l|l||}
\hline
\textbf{SE NFR (Canonical)} & \textbf{PROMISE Types Mapped} \\
\hline
\hline
Functional & F \\
Security & SE \\
Usability (incl. Look-and-feel) & US, LF \\
Operability / Supportability & O \\
Performance & PE \\
Availability & A \\
Maintainability & MN \\
Scalability & SC \\
Reliability (incl. Fault-Tolerance) & FT \\
Legal / Compliance & L \\
Portability & PO \\
\hline
\hline
\end{tabular}
\caption{Aligning PROMISE\_exp requirements with systems-engineering non-functional taxonomy.}
\label{tab:promise_se_nfr_mapping}
\end{table*}

Table \ref{tab:req_type_distribution} (Original Promise\_ext break down), provided fundamental insights into the distribution of requirement types used in the study. This data served as the bedrock for evaluating the performance of the artificial intelligence models in classifying software requirements as Functional (FR) versus Non-Functional (NFR). 

\begin{table*}[ht]
\centering
\begin{tabular}{||l |c |r||}
\hline
\textbf{Requirement Type} & \textbf{Code} & \textbf{Count} \\
\hline
\hline
Functional Requirement & F  & 444 \\
Availability            & A  & 31  \\
Legal                   & L  & 15  \\
Look-and-feel           & LF & 49  \\
Maintainability         & MN & 24  \\
Operability             & O  & 77  \\
Performance             & PE & 67  \\
Scalability              & SC & 22  \\
Security                & SE & 125 \\
Usability               & US & 85  \\
Fault Tolerance         & FT & 18  \\
Portability             & PO & 12  \\
\hline
\hline
\textbf{Total}           &     & \textbf{969} \\
\hline
\hline
\end{tabular}
\caption{Original Promise\_ext break down}
\label{tab:req_type_distribution}
\end{table*}

\begin{figure}[H]
  \centering
  \colfig{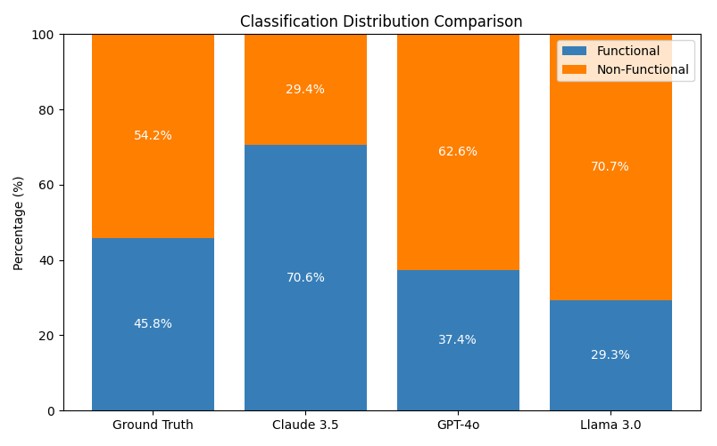}
  \caption{FR vs. NFR Prediction Distribution}
  \label{Fig:Fig9}
\end{figure}

\mysubsection{Classification Distribution.}
Figure \ref{Fig:Fig9} illustrates how different models compare to the reference classification by showing the proportion of items each one assigns to the two categories. It highlights the overall tendency of each model and allows a visual comparison of how their distributions align with or diverge from the reference pattern.
It summarizes the comparative distribution of functional and non-functional classifications across the ground truth and three large language models.

As we can observe in figure 7 the original data base exhibits a relatively balanced distribution, with 45.8\% of the requirements labeled as Functional and 54.2\% as Non-Functional. In contrast, Claude 3.5 displays a strong functional bias, assigning 70.6\% of the items to the Functional category and only 29.4\% to the Non-Functional category. GPT-4o shows a milder deviation, producing 37.4\% Functional and 62.6\% Non-Functional classifications, indicating a moderate inclination toward Non-Functional labeling. Llama 3.0 diverges the most from the ground truth, with 29.3\% classified as Functional and 70.7\% as Non-Functional, reflecting a pronounced non-functional bias.
Collectively, these discrepancies highlight systematic tendencies within each model, demonstrating that none of them accurately replicates the empirical distribution. Each model exhibits a distinct directional bias—functional overestimation in Claude and non-functional overestimation in GPT-4o and Llama—indicating differing sensitivities to linguistic cues present in the dataset.

\mysubsection{Model Performance Observations.}
The following points highlight behavior differences identified between models, addressing recall differences, subtype trends, and category-specific patterns. Recall measures a model’s ability to correctly identify all relevant instances of a given class, defined as the proportion of true positives among all actual positives, reflecting how many relevant instances are successfully detected versus missed. These results are summarized in Table 5.

\paragraph{Functional Requirements ($n=444$):}
\begin{itemize}
    \item \emph{Llama~3.0} achieved the highest recall at 86.3\%, representing a 41.7-point advantage over Claude. This performance is attributed to effective pattern matching when ``shall\ldots'' constructs are used in requirement statements.
    \item \emph{Claude} exhibited low recall (44.6\%), indicating a bias toward labeling requirements as non-functional.
\end{itemize}

\paragraph{Non-Functional Requirement Subcategories ($n=525$):}
\begin{itemize}
    \item \emph{Claude~3.5~Sonnet} dominated all NFR subtypes, with particularly strong performance in \textit{Maintainability} (91.7\%) and \textit{Fault Tolerance} (88.9\%), reflecting its strength in detecting nuanced attribute-oriented language.
    \item The largest performance gaps were observed in less frequent categories (FT, MN, LF), where specialized terminology and smaller sample sizes amplify inter-model differences.
    \item Perfect scores in \textit{Portability} (L and PO) indicate consistent keyword cues that the model reliably captures.
\end{itemize}

This study demonstrates that large language models can effectively automate the distinction between functional requirements (FR) and non-functional requirements (NFR); however, performance varies by task formulation and prompting strategy. Specifically:

\begin{itemize}
    \item FR versus NFR classification is best handled by \emph{GPT-4o} (approximately 85\% accuracy) when using a clear, single-label prompt (``Classify as Functional or Non-Functional'').
    \item NFR subcategory assignment benefits from models specialized in attribute-based language; \emph{Claude~3.5~Sonnet} consistently outperformed other models across quality, performance, security, and reliability categories.
    \item Simple ``shall\ldots'' statements representing pure functional requirements are reliably recognized by \emph{Llama~3.0}; however, it tends to under-detect NFRs without additional prompting.
\end{itemize}

\begin{table*}[ht]
\centering
\small
\begin{tabular}{||l |c |c |p{5.2cm}||}
\hline
\hline
\textbf{Model} & \textbf{Task} & \textbf{Recall (\%)} & \textbf{Key Observations} \\
\hline
Llama~3.0 &
Functional Requirements (FR) &
86.3\% &
Highest recall among all models; demonstrates strong pattern matching, especially for explicit ``shall''-based requirement formulations. \\

GPT 4.0 &
Functional Requirements (FR) &
44.6\% &
Low recall due to systematic bias toward labeling requirements as non-functional, even when functional intent is explicit. \\

Claude~3.5~Sonnet &
Non-Functional Requirements (NFR) &
$>$88\% (up to 91.7\%) &
Dominates all NFR subcategories; excels in Maintainability (91.7\%) and Fault Tolerance (88.9\%) due to sensitivity to nuanced, attribute-oriented language. \\

\hline
\hline
\end{tabular}
\caption{Comparative recall performance across models for Functional and Non-Functional Requirement classification.}
\label{tab:fr_nfr_recall}
\end{table*}

Claude 3.5 processed all 969 requirements correctly. GPT-4o and Llama 3.0 produced ~12 \% duplicate or split entries.

\section{Conclusions and Contributions}

This research examined the use of Large Language Models (LLMs) as AI “copilots” in requirements engineering, with a focus on their ability to support requirement quality assessment and classification within an INCOSE-aligned framework. The findings demonstrate that AI can assume a portion of routine analytical work, provided its deployment is embedded within a structured human-in-the-loop (HITL) process.

Regarding \emph{RQ1}, the study shows that modern LLMs can classify requirements according to INCOSE “good requirement” criteria with accuracy comparable to experienced systems engineers. High alignment was observed in identifying problematic requirements, particularly for structural and linguistic defects. However, a consistent \textit{rationale gap} was identified: while AI and human experts often agree on the need for revision, they may differ on the underlying cause. AI emphasizes formal rule violations, whereas human experts prioritize ambiguity, missing context, and conceptual intent.

For \emph{RQ2}, the analysis confirms that AI can distinguish between functional and non-functional requirements by identifying underlying intent rather than relying solely on syntax. Different models exhibit distinct classification biases, indicating that model selection must align with project priorities, especially in safety-critical or performance-driven systems.

In addressing \emph{RQ3}, the results highlight AI’s primary advantages in efficiency, consistency, and repeatability. AI applies evaluation criteria uniformly and does not suffer from cognitive fatigue or subjective drift. Conversely, AI’s limitations remain noticeable at this stage of the technology: it does not provide totally accurate performance with regard to technical truth, physical feasibility, and cross-document engineering logic, and remains susceptible to hallucinations. 

These gaps between AI capabilities and human expertise directly inform how AI should be operationalized. Rather than viewing AI's limitations as a barrier, they define a clear boundary for a \emph{workflow shift} in requirements engineering. By delegating structural and linguistic auditing to AI, systems engineers can redirect cognitive effort toward high-value validation activities, including feasibility assessment, trade-off analysis, and architectural reasoning.

To operationalize this shift, we propose a concrete, AI-assisted workflow allocation. Table \ref{tab:decision_workflow} maps the specific responsibilities of the AI copilot versus the human expert across the seven evaluated INCOSE criteria.

\begin{table*}[ht]
\centering
\small
\begin{tabular}{||l | p{6.5cm} | p{6.5cm}||}
\hline
\textbf{INCOSE Criterion} & \textbf{AI Copilot Role (Pre-Audit)} & \textbf{Human Expert Role (Validation \& Decision)} \\
\hline
\hline
\emph{Unambiguous} & Flags vague terminology and linguistic inconsistencies. & Resolves domain-specific ambiguity and contextual nuances. \\
\emph{Singular} & Identifies and splits compound requirements (e.g., detecting "and/or"). & Reviews and approves the separated requirement statements. \\
\emph{Independent} & Detects implementation-specific language and design bias. & Ensures true solution neutrality based on system architecture. \\
\emph{Complete} & Highlights missing parameters, units, or standard constraints. & Validates conceptual completeness against stakeholder needs. \\
\emph{Verifiable} & Checks for quantifiable metrics and measurable targets. & Confirms actual testability within project resources and constraints. \\
\rowcolor{tableheader}
\emph{Essential} & \textit{Limited.} Flags potential duplicates or out-of-scope keywords. & \emph{Primary Driver.} Determines necessity for the system's core mission. \\
\rowcolor{tableheader}
\emph{Feasible} & \textit{Limited.} Flags extreme metric anomalies based on training data. & \emph{Primary Driver.} Assesses technical, budgetary, and schedule realism. \\
\hline
\hline
\end{tabular}
\caption{Decision Allocation: Mapping AI vs. Human Responsibilities across INCOSE Quality Criteria.}
\label{tab:decision_workflow}
\end{table*}

Based on this allocation, a recommended three-step workflow emerges for industrial practice:
\begin{enumerate}
    \item \emph{Initial AI Pre-Audit:} The LLM autonomously scans draft requirements, resolving Singular/Unambiguous violations and tagging issues related to Verifiability and Independence.
    \item \emph{Human Review \& Reconciliation:} The systems engineer reviews AI-flagged items, accepting or modifying the structural recommendations and resolving contextual nuances.
    \item \emph{Expert Validation:} Relieved of basic syntax checking, the engineer focuses deeply on the Essential and Feasible criteria, applying domain expertise and architectural judgment.
\end{enumerate}

This approach demonstrates that LLM-based tools can perform structural and linguistic analysis in alignment with INCOSE guidelines, achieving near expert-level agreement for routine tasks. By introducing AI as a standardized pre-audit layer, organizations can reduce early lifecycle review effort and peer-review bottlenecks, while strictly preserving expert authority over system validation and decision-making.

\subsection{Future Applications and Potential Challenges}
AI presents significant opportunities for RE that can be expected, including automated requirements elicitation from stakeholder interviews, synthesis of requirements from high-level goals, personalization of documentation for diverse audiences, and continuous monitoring for evolving needs. Large language models can already draft candidate requirements, detect ambiguities, and propose refinements and are continuously evolving. Integration of AI with alternative media such as visual models could enhance clarity and stakeholder communication (\cite{Kaur2024}).
Key challenges remain: ensuring validation and trust in AI outputs, establishing accountability for AI-generated requirements, safeguarding sensitive data, maintaining ethical and regulatory compliance, and integrating AI tools into the inherently social RE process. Training engineers to use AI effectively is also critical, especially in the range of possibilities for automation, but also notably in the limitations as described in this research.
There is a need for further empirical evidence on real-world adoption, frameworks for responsible AI use, and multidisciplinary collaboration. With careful governance, AI combined with established RE practices is starting to show a potential to improve efficiency, manage complexity, and enhance requirement quality, ultimately leading to better systems, but as noted, more empirical research is required.

\section{Acknowledgments}
The conceptual design of the research, its implementation, analysis of results and conclusions are original authors' work. ChatGPT 5.2 assisted with consistency checking of the structure and flow of the paper, assisting in identifying repetitive paragraphs, and English corrections. It assisted in a limited way in the generation of the paper's abstract which was extensively revised by the authors.

\section{References}
\printbibliography[heading=none]

@misc{Stanford2024,
  author       = {{AI Index Steering Committee}},
  title        = {{AI} Index 2024 Annual Report},
  year         = {2024},
  howpublished = {\url{https://aiindex.stanford.edu/wp-content/uploads/2024/05/HAI_AI-Index-Report-2024.pdf}},
  institution  = {Stanford Institute for Human-Centered Artificial Intelligence},
  address      = {Stanford, CA}
}

@manual{nasa2016_seh,
  author       = {{NASA}},
  title        = {NASA Systems Engineering Handbook, Chapter 2.0: Fundamentals of Systems Engineering},
  year         = {2016},
  organization = {National Aeronautics and Space Administration},
  note         = {NASA/SP-2016-6105 Rev 2},
  url          = {https://www.nasa.gov/reference/2-0-fundamentals-of-systems-engineering/}
}

@inproceedings{Baidya2022DigitalTwin,
  author    = {Sabur Baidya and Sumit K. Das and Mohammad Helal Uddin and Chase Kosek and Chris Summers},
  title     = {Digital Twin in Safety-Critical Robotics Applications: Opportunities and Challenges},
  booktitle = {Proceedings of the 2022 {IEEE} International Performance, Computing, and Communications Conference ({IPCCC})},
  pages     = {101--107},
  year      = {2022},
  publisher = {IEEE},
  doi       = {10.1109/IPCCC55026.2022.9894313}
}

@misc{TeslaAIPage,
  author       = {{Tesla}},
  title        = {Artificial Intelligence at {Tesla}},
  howpublished = {\url{https://www.tesla.com/AI}},
  year         = {2025}
}

@article{Cheligeer2022,
  author  = {Chinbat Cheligeer and Jun Huang and Guangyu Wu and Nayeem Bhuiyan and Yong Xu and Yuping Zeng},
  title   = {Machine Learning in Requirements Elicitation: A Literature Review},
  journal = {{AI} EDAM},
  volume  = {36},
  pages   = {e32},
  year    = {2022},
  doi     = {10.1017/S0890060422000103}
}

@inproceedings{TamaiAnzai2018,
  author    = {Tetsuo Tamai and Taichi Anzai},
  title     = {Quality Requirements Analysis with Machine Learning},
  booktitle = {Proceedings of the 13th International Conference on Evaluation of Novel Approaches to Software Engineering ({ENASE})},
  pages     = {241--248},
  year      = {2018}
}

@manual{INCOSE2023,
  author       = {{INCOSE}},
  title        = {Guide to Writing Requirements (Version 4)},
  organization = {International Council on Systems Engineering},
  year         = {2023},
  url          = {https://www.incose.org/docs/default-source/working-groups/requirements-wg/gtwr/incose_rwg_gtwr_v4_040423_final_drafts.pdf?sfvrsn=5c877fc7_2}
}

@inproceedings{Maleki2024,
  author    = {Niloofar Maleki and Pranav Soni and Karthik Padmanabhan and Kumar Dutta},
  title     = {{AI} Hallucinations: A Misnomer Worth Clarifying},
  booktitle = {2024 {IEEE} Conference on Artificial Intelligence ({CAI})},
  pages     = {133--138},
  year      = {2024},
  publisher = {IEEE},
  doi       = {10.1109/CAI58096.2024.00026}
}

@article{Martinez2023,
  author  = {Gustavo Mart{\'i}nez and Jorge Conde and Pedro Reviriego and Eduardo Merino-G{\'o}mez and Jos{\'e} Antonio Hern{\'a}ndez and Fabrizio Lombardi},
  title   = {How Many Words Does {ChatGPT} Know? The Answer is {ChatWords}},
  journal = {arXiv preprint arXiv:2309.16777},
  year    = {2023},
  doi     = {10.48550/arXiv.2309.16777}
}

@misc{Hadar2022,
  author        = {Hadar, A. and Levy, N. and Winokur, M.},
  title         = {Management and Detection System for Medical Surgical Equipment},
  year          = {2022},
  eprint        = {2211.02351},
  archivePrefix = {arXiv},
  url           = {https://arxiv.org/abs/2211.02351}
}

@article{PerezCerrolaza2024,
  author  = {Javier P{\'e}rez-Cerrolaza and Javier Abella and Marcus Borg and Carlo Donzella and Jes{\'u}s Cerquides and Francisco J. Cazorla and Juan L. Flores},
  title   = {Artificial Intelligence for Safety-Critical Systems in Industrial and Transportation Domains: A Survey},
  journal = {{ACM} Computing Surveys},
  volume  = {56},
  number  = {7},
  pages   = {1--40},
  year    = {2024},
  doi     = {10.1145/3626314}
}

@article{Siddique2022,
  author  = {Iqtiar M. Siddique},
  title   = {Systems Engineering in Complex Systems: Challenges and Strategies for Success},
  journal = {European Journal of Advances in Engineering and Technology},
  volume  = {9},
  number  = {9},
  pages   = {61--66},
  year    = {2022},
  doi     = {10.5281/zenodo.11545350}
}

@article{Kolligs2025,
  author  = {Kolligs, Johannes W. and Thomas, Lucas D. W.},
  title   = {Characterizing Efficacy of Alternative Media for Requirements Expression},
  journal = {Systems},
  year    = {2025},
  volume  = {13},
  number  = {5},
  pages   = {314},
  doi     = {10.3390/systems13050314}
}

@article{Dalpiaz2020,
  author  = {Dalpiaz, Fabiano and Niu, Narjes},
  title   = {Requirements Engineering in the Days of Artificial Intelligence},
  journal = {{IEEE} Software},
  year    = {2020},
  volume  = {38},
  number  = {4},
  pages   = {7--12},
  doi     = {10.1109/MS.2020.2973364}
}

@article{Siddique2022b,
  author  = {Siddique, Imtiaz Mohammed},
  title   = {Harnessing Artificial Intelligence for Systems Engineering: Promises and Pitfalls},
  journal = {European Journal of Advances in Engineering and Technology},
  year    = {2022},
  volume  = {9},
  number  = {9},
  pages   = {67--72},
  doi     = {10.5281/zenodo.11545453}
}

@article{DalpiazNiu2020,
  author  = {Fabiano Dalpiaz and Nan Niu},
  title   = {Requirements Engineering in the Days of Artificial Intelligence},
  journal = {{IEEE} Software},
  volume  = {37},
  number  = {4},
  pages   = {7--10},
  year    = {2020},
  doi     = {10.1109/MS.2020.2986047}
}

@article{Kaur2024,
  author  = {Komalpreet Kaur and Preeti Kaur},
  title   = {The Application of {AI} Techniques in Requirements Classification: A Systematic Mapping},
  journal = {Artificial Intelligence Review},
  volume  = {57},
  pages   = {Article 57},
  year    = {2024},
  doi     = {10.1007/s10462-023-10667-1}
}

@article{Vaswani2017,
  author  = {Ashish Vaswani and Noam Shazeer and Niki Parmar and Jakob Uszkoreit and Llion Jones and Aidan N. Gomez and {\L}ukasz Kaiser and Illia Polosukhin},
  title   = {Attention Is All You Need},
  journal = {Advances in Neural Information Processing Systems},
  volume  = {30},
  year    = {2017},
  url     = {https://arxiv.org/abs/1706.03762}
}

@article{MartinezFernandez2022,
  author  = {Mart{\'i}nez-Fern{\'a}ndez, Silvia and Bogner, Justus and Franch, Xavier and Oriol, Marc and Siebert, Julian and Trendowicz, Adam and others},
  title   = {Software Engineering for {AI}-Based Systems: A Survey},
  journal = {{ACM} Transactions on Software Engineering and Methodology},
  year    = {2022},
  volume  = {31},
  number  = {4},
  article-number = {76},
  doi     = {10.1145/3557944}
}

@inproceedings{Peer2024,
  author    = {Jordan Peer and Yaniv Mordecai and Yoram Reich},
  title     = {{NLP4ReF}: Requirements Classification and Forecasting:
               From Model-Based Design to Large Language Models},
  booktitle = {2024 {IEEE} Aerospace Conference},
  pages     = {1--16},
  year      = {2024},
  publisher = {{IEEE}},
  doi       = {10.1109/AERO58547.2024.10521022},
  url       = {https://ieeexplore.ieee.org/document/10521022}
}

@inproceedings{Bender2021,
  author    = {Emily M. Bender and Timnit Gebru and Angelina McMillan-Major and Shmargaret Shmitchell},
  title     = {On the Dangers of Stochastic Parrots: Can Language Models Be Too Big?},
  booktitle = {Proceedings of the 2021 {ACM} Conference on Fairness, Accountability, and Transparency ({FAccT} '21)},
  pages     = {610--623},
  year      = {2021},
  publisher = {ACM}
}

@article{Shneiderman2020,
  author  = {Ben Shneiderman},
  title   = {Human-Centered Artificial Intelligence: Reliable, Safe \& Trustworthy},
  journal = {International Journal of Human-Computer Interaction},
  volume  = {36},
  number  = {6},
  pages   = {495--504},
  year    = {2020}
}

@inproceedings{ClelandHuang2006,
  author    = {Jane Cleland-Huang and Robert Settimi and Xuchang Zou and Peter Solc},
  title     = {The Detection and Classification of Non-Functional Requirements},
  booktitle = {Proceedings of the 14th {IEEE} International Requirements Engineering 
               Conference ({RE}'06)},
  pages     = {39--48},
  year      = {2006},
  doi       = {10.1109/RE.2006.65}
}

\newpage
\phantomsection
\makeatletter
\renewcommand{\authorbioentry}[3]{%
  \noindent\begin{tabular}{@{}m{0.5in} M{\dimexpr\columnwidth-0.5in\relax}@{}}
    \authorpic{#1} &
    {\headingfont\bfseries\raggedright\fontsize{12pt}{14pt}\selectfont #2}\par #3
  \end{tabular}\par\medskip
}
\makeatother


\end{multicols*}

\end{document}